# Body-worn triaxial accelerometer coherence and reliability related to static posturography in unilateral vestibular failure

*Impiego degli accelerometri triassiali nel deficit vestibolare unilaterale: affidabilità rispetto alla posturografia statica*

M. ALESSANDRINI[1], A. MICARELLI[1,2], A. VIZIANO[1], I. PAVONE[1,3], G. COSTANTINI[4], D. CASALI[4], F. PAOLIZZO[5], G. SAGGIO[4]

[1] Department of Clinical Sciences and Translational Medicine, University of Rome Tor Vergata, Italy; [2] Department of Systems Medicine, Neuroscience Unit, University of Rome Tor Vergata, Italy; [3] Otolaryngology and Head and Neck Surgery Unit, "Santo Spirito" Hospital of Pescara, Italy; [4] Department of Electronic Engineering, University of Rome Tor Vergata, Italy; [5] Department of Cognitive Sciences, University of California, Irvine, USA

SUMMARY

Since changes in vestibular function may be one cause of disequilibrium, major advances in measuring postural control and sensory integration in vestibular impairments have been achieved by using posturography. However, in order to overcome problems related to this type of technology, body-worn accelerometers (ACC) have been proposed as a portable, low-cost alternative to posturography for measurements of postural sway in a friendly and ecologic environment. Due to the fact that no study to date has shown the experimental validity of ACC-based measures of body sway with respect to posturography for subjects with vestibular deficits, the aim of the present study was: i) to develop and validate a practical tool that can allow clinicians to measure postural sway derangements in an otoneurological setting by ACC, and ii) to provide reliable, sensitive and accurate automatic analysis of sway that could help in discriminating unilateral vestibular failure (UVF) patients. Thus, a group of 13 patients (seven females, 6 males; mean age 48.6 ± 6.4 years) affected for at least 6 months by UVF and 13 matched healthy subjects were instructed to maintain an upright position during a static forceplate-based posturography (FBP) acquisition while wearing a Movit® sensor (by Captiks) with 3-D accelerometers mounted on the posterior trunk near the body centre of mass. Pearson product moment correlation demonstrated a high level of correspondence of four time-domain and three frequency-domain measures extracted by ACC and FBP testing; in addition, t-test demonstrated that two ACC-based time- and frequency-domain parameters were reliable measures in discriminating UVF subjects. These aspects, overall, should further highlight the attention of clinicians and researchers to this kind of sway recording technique in the field of otoneurological disorders by considering the possibility to enrich the amount of quantitative and qualitative information useful for discrimination, diagnosis and treatment of UVF. In conclusion, we believe the present ACC-based measurement of sway offers a patient-friendly, reliable, inexpensive and efficient alternative recording technique that is useful – together with clinical balance and mobility tests – in various circumstances, as well as in outcome studies involving diagnosis, follow-up and rehabilitation of UVF patients.

RIASSUNTO

*Poichè le alterazioni della funzione vestibolare possono essere causa di disequilibrio, i principali reperti sviluppati ad oggi per misurare il controllo posturale e l'integrazione sensoriale nel danno vestibolare sono stati ottenuti grazie alla posturografia. Tuttavia, al fine di superare i problemi legati a tale genere di tecnologia, sono stati proposti gli accelerometri indossabili (ACC) come un'alternativa portatile e a basso costo per la misurazione dell'oscillazione corporea in ambienti confortevoli. D'altro canto, nessuno studio ad oggi ha dimostrato la validità sperimentale delle misura- zioni ottenute con ACC - rispetto a quelle derivanti dalla posturografia - in soggetti affetti da deficit vestibolare. Pertanto, l'obiettivo del presente lavoro è stato quello di i) sviluppare e validare una strumentazione pratica che potesse consentire la misurazione dei disordini dell'oscillazione corporea nell'ambito della valutazione otoneurologica attraverso gli ACC e ii) fornire un'analisi delle oscillazioni affidabile ed automatica, che potesse implementare in modo sensibile ed accurato la possibile discriminazione di pazienti affetti da deficit vestibolare unilaterale (UVF). A tale scopo, un gruppo di 13 pazienti (sette femmine, 6 maschi; età media 48.6 ± 6.4 anni) affetti da UVF da almeno 6 mesi e un altro omogeneo di 13 soggetti sani sono stati invitati a mantenere la posizione eretta durante l'esecuzione della posturografia statica (FBP) mentre indossavano a livello lombare - vicino al centro di massa - un sensore Movit® (by Captiks) costituito da accelerometri 3-D. La correlazione 'product-moment' secondo Pearson ha dimostrato un elevato livello di corrispondenza di quattro misure, estratte da ACC e da FBP, nel dominio del tempo e di tre in quello della frequenza. Inoltre il t-test ha evidenziato che due parametri nel dominio del tempo e due in quello della frequenza si sono dimostrati affidabili nel discriminare i soggetti affetti da UVF. Tali aspetti, nel loro complesso, dovrebbero focalizzare l'attenzione in ambito clinico e di ricerca su tale tecnica di registrazione, considerato l'arricchimento quantitativo e qualitativo di informazioni utili nella discriminazione, diagnosi e trattamento di pazienti affetti da UVF. In conclusione, noi riteniamo che la misurazione basata su ACC offra un'alternativa confortevole, affidabile, economica ed efficiente utile, assieme ai test clinici di equilibrio e mobilità, in molteplici circostanze così come negli studi implicati nella diagnosi, controllo e riabilitazione di pazienti affetti da UVF.*

PAROLE CHIAVE: *Accelerometro • Deficit vestibolare unilaterale • Posturografia statica • Video test impulsivo del capo*

## Introduction

Among the general population, 20-30% of individuals experience balance disorders that affect daily activities [1,2] inducing, as a consequence, a higher risk of bodily fall, which is becoming a serious social problem, especially for the elderly [1,2]. As is known, appropriate equilibrium in space depends on the integration of vision, proprioception and vestibular information [1]. These afferent sensory feedback signals play a crucial role in adapting and modulating the operation of the locomotor and stance network in the real environment [1]. Considering also that changes in vestibular function may be one cause of ataxia, major findings in measurement of postural control and sensory integration in vestibular impairments were achieved using static posturography in an ambulatory or forced-experimental setting [3].

This type of technology currently uses force plate analysis of centre of pressure (COP) displacement during quiet stance, the sensitivity of which has been demonstrated to be reliable to measure postural disorders in neurological, systemic and vestibular diseases and in fall risk in the elderly [3,4].

However, forceplate-based posturography (FBP; or stabilometry) is a large and expensive instrumentation that requires proper installation, and as such may not be practical for clinical use [1,4]. Recently, in order to overcome these problems, body-worn accelerometers (ACC) have been proposed as a portable, low-cost alternative to a force plate for measurements of postural sway in a friendly and ecologic environment [1,2,4,5]. Despite the potential advantages of accelerometric systems in clinical practice, they still have several drawbacks, such as the need to pre-process data and the question of how to translate sway measures into clinically-understandable outcomes [4-6]. However, the major limitation is that there is no consensus as to which sway-related measures should be considered [4-6]. In fact, in order to make ACC-based measures useful for clinical applications, it is important to assess their validity, sensitivity and reliability compared to gold standard laboratory and clinical assessments [4,6], such as FBP, in vestibular clinical and research. The relationship between the same postural sway measures calculated from the force plate COP and from ACC that would support the experimental validity of ACC measures has only been reported in few studies, and none have demonstrated the experimental validity of ACC-based and force plate-based measures of postural sway for subjects with vestibular deficits [4,6].

The aim of the present study was to develop and validate a practical tool that could allow clinicians to measure postural sway derangements in an otoneurological setting using body-worn ACC. Our vision is that this tool will provide reliable, automatic analysis of sway that is sensitive, accurate, robust and consistent, without the need for clinical experts to deal with raw data. To determine if the sensitivity and experimental concurrent validity of ACC – compared to FBP measures of postural sway – could be a reliable tool in diagnosing and discriminating vestibular impairments, a group of unilateral vestibular failure (UVF) and age-, body mass index (BMI) and gender- matched healthy subjects (HS) were enrolled.

## Materials and methods

*Participants*

Thirteen subjects (seven females, 6 males; mean age 48.6 ± 6.4 years; body mass index, BMI = 22.3 ± 2.1 kg/m$^2$) affected for at least 6 months by UVF and 13 BMI-, gender- and age-matched HS (6 females, 7 males; mean age 47.7 ± 6.1 years; BMI = 21.9 ± 2.3 kg/m$^2$) participated in the study. After thorough clinical otoneurological examination (binocular electrooculography analysis, Head Shaking Test, clinical Head Impulse Test as well as limb coordination, gait observation and Romberg stance Test) all subjects underwent video Head Impulse Test (vHIT) measurements using the EyeSeeCam™ System and the technique proposed by Blodow et al. [7]. The vHIT results were defined as normal if they were within the calculated gain-reference range, mean$_{normal}$ ± 2 standard deviations (SD), incorporating 95% of population and if no refixation saccades occurred. For the HS, the average VOR gain was 0.97 ± 0.12 (mean$_{normal}$ ± SD) on the right side and 0.98 ± 0.14 on the left side. A t-test was performed between both sides VOR gain and no significant difference ($p < 0.05$) was found. Thus, diagnosis of UVF was achieved in case of values under a gain threshold set to 0.73 for the right side and to 0.7 for the left side. Right and left UVF subjects demonstrated an average VOR gain of 0.44 ± 0.18 and 0.42 ± 0.19, respectively.

The protocol adhered to the principles of the Declaration of Helsinki and all participants provided written informed consent after receiving a detailed explanation of the study.

*Static posturography testing*

Each subject was instructed to maintain an upright position on a standardised platform for static posturography (EDM Euroclinic®). The recording period was 60 sec for each test (eyes closed [CE] or opened [OE] while standing on the stiff platform [SP] or a 6 cm heightened foam carpet [FC]); the sampling frequency was 25 Hz [3]. The COP was monitored while performing the test.

*Accelerometer implementation*

According to previous studies [4], during all conditions subjects wore an inertial measurement unit termed Movit® (by Captiks Srl, Italy), equipped with 3-D accelerometers (±1.7 g range) mounted on the posterior trunk at the level of L5 by means of an elastic belt, near the body centre of mass (Fig. 1). The sensing axes were oriented along the anatomical antero-posterior (AP), medio-lateral (ML), and vertical directions.

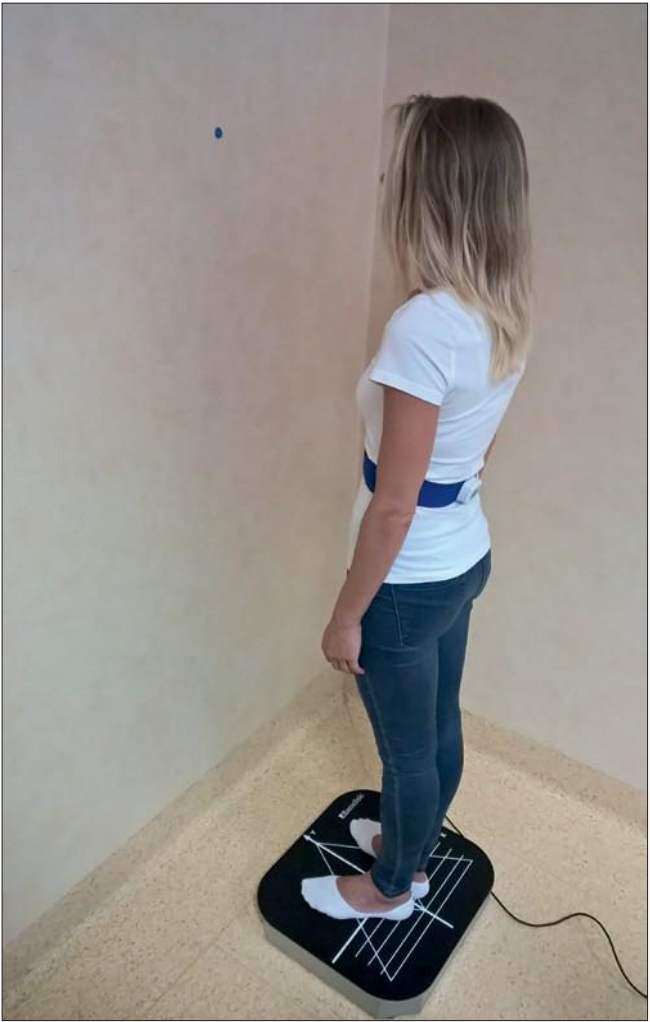

Fig. 1. Experimental setting and body-worn accelerometer position.

The Movit® combines a 3D accelerometer with a 3D gyroscope, a 3D compass and a barometer forming a complete position and orientation tracker. Combination of sensors allows data integration to reduce drift and/or cumulative measurement errors. All sensors were handled by an AT32UC3A4256 microcontroller (by Atmel, San Jose, California, US), with a frequency clock of 40 MHz. The Movit® uses an 802.15.4 protocol to send wireless data to a receiving base station, which feds data to a personal computer at a selectable data rate within 4 Hz-1000 Hz. The power supply has a USB-rechargeable built-in Li-Po battery that can provide up to eight hours of operation.

In particular, we focused on data coming from the 3D accelerometer, which integrates a 16-bit A/D converter, and allows options of ±2 g, ±4 g, ±8 g and ±16 g full scale acceleration range. For our purposes, the choice was the ±16 g option, and the low-power mode of the Movit® was selected. corresponding at an operative current of 140 μA at a 50 Hz updating data rate. For our purposes, we considered two out of three parameters of the accelerometer, namely the AP (or pitch) angle, and the ML (or roll) angle.

The receiving personal computer was equipped with 4 GB of RAM and an i5 processor (by Intel, Santa Clara, California, US), and data were elaborated using ad-hoc written software routines in Matlab (by Matworks, Natick, Massachusetts, US).

*Data handling and statistical analysis*

After a MATLAB-based automatic check of normality distribution of parameters, a total of 11 ACC measures previously validated in literature [4] were computed from the 2D acceleration time series, similar to COP analysis. Following the same study [4] and details reported in Table I, for each trial we computed in the time-domain six measures that characterised the ACC trajectory: mean distance from centre of COP trajectory (DIST), root mean square of COP time series (RMS), total length of COP trajec- tory (PATH), range of COP displacement or acceleration (RANGE), the number of loops that have to be run by the COP to cover a total trajectory equal to PATH (MF) and the sway area computed as the area spanned from the COP per unit of time (AREA). In the frequency-domain, spectral properties were assessed for each trial by five measures: one that quantifies the total power of the signal (PWR), one that estimates the variability of the frequency content of the power spectral density (FD) and three measures of characteristic frequencies in the power spectral density (median frequency, below which the 50% of PWR is present, F50; 95% power frequency, below which the 95% of PWR is present, F95; and the centroidal frequency, CF) (Table I).

Table I. Summary of the extracted measures. Accelerometer, ACC; centre of pressure, COP.

| Measure | Description |
|---|---|
| DIST | Mean distance from centre of COP (ACC) trajectory [mm] ([m/s$^2$]) |
| RMS | Root mean square of COP (ACC) time series [mm] ([m/s$^2$]) |
| PATH | Sway path, total length of COP (ACC) trajectory [mm] ([m/s$^2$]) |
| RANGE | Range of COP displacement (acceleration) [mm] ([m/s$^2$]) |
| MF | Mean frequency, the number, per second, of loops that have to be run by the COP (ACC), to cover a total trajectory equal to PATH (MF = PATH/(2*π*DIST*trial duration) (Hz) |
| AREA | Sway area, computed as the area spanned from the COP (ACC) per unit of time [mm$^2$/s] ([m$^2$/s$^5$]) |
| PWR | Total power [mm$^2$] ([m$^2$/s$^4$]) |
| F50 | Median frequency, frequency below which the 50% of PWR is present (Hz) |
| F95 | 95% power frequency, frequency below which the 95% of PWR is present (Hz) |
| CF | Centroidal frequency (Hz) |
| FD | Frequency dispersion (−) |

Algorithms for signal analysis and statistical evaluation of outcomes were written in MATLAB.

A Pearson product moment correlation was used to assess the relationship between COP and ACC metrics. Differences between UVF and control groups were determined using a t-test. Differences were assumed significant when $p < 0.05$.

## Results

The major results of this technical report are represented by a significant correlation both in UVF and HS subjects in PATH, RANGE, MF, AREA, F50, F95 and CF (Table II, Fig. 2). Moreover, a statistically significant difference was found between UVF and HS subjects in AREA, MF, F50 and F95 parameter scores in many of tested conditions (for details see Table II).

## Discussion

In the present study we tested, for the first time, validated ACC-based time-domain and frequency-domain measures in a group of UVF patients with the attempt of defining their possible usefulness, with regard to FBP, in discriminating vestibular impairments.

Sensors measures of postural sway were validated by force-plate measures of COP displacement and many, but not all, ACC-based measures were correlated with the gold-standard laboratory measures of sway from a force-plate. In particular, the first interesting finding was represented by the high level of robust correspondence of four time-domain (PATH, RANGE, MF, AREA) and three frequency-domain (F50, F95 and CF) measures extracted by ACC and FBP testing, when recorded in same subjects during same conditions (Table II). According to previous studies, if the body was thought to be moving like an inverted pendulum, a correlation close to 1 would be expected between trunk acceleration and COP displacement [4]. Thus, highly correlated measures of COP and centre of mass amplitudes have also been reported during quiet stance [4]. However, a possible explanation of the absence of correlation in some measures could be related to the fact that subjects do not sway strictly as inverted pendulums [4]. In fact, even quiet stance in young, healthy subjects includes some hip strategy, and the amount of hip strategy used to control posture have been shown to increase with age [3 4]. Besides this, to the best of our knowledge, in the present work the reliability of important postural sway measures in UVF subjects using an accelerometer-based approach is reported.

Secondly, when comparing parameters extracted in all conditions by ACC-based and COP-based recordings two important time-domain measures – MV and AREA – were further demonstrated to be significant reliable parameters

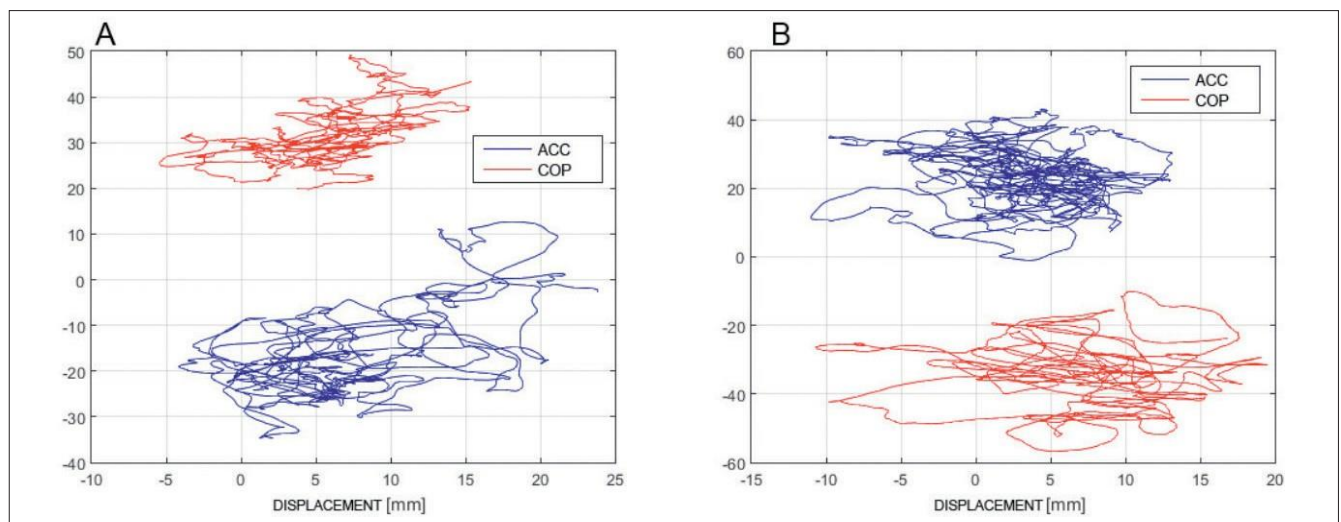

Fig. 2. Accelerometer (ACC) and centre of pressure (COP) traces in the horizontal plane during opened eyes stance on a stiff platform for a representative healthy (A) and unilateral vestibular failure (B) subject.

Table II. Sensitivity of COP and ACC-based measures.

| | | COP | | | | | ACC | | | | | Correlation | |
|---|---|---|---|---|---|---|---|---|---|---|---|---|---|
| | | UVF | | HS | | T-test | UVF | | HS | | T-test | | |
| | | MEAN | SD | MEAN | SD | p values | MEAN | SD | MEAN | SD | p values | r | P |
| **DIST** | SP-OE | 6.8 | 1.43 | 5.4 | 1.58 | 0.1046 | 6.78 | 2.19 | 8.72 | 2.51 | 0.2297 | -0.05 | 0.32838 |
| | SP-CE | 7.72 | 2.35 | 6.96 | 2.17 | 0.2774 | 6.72 | 1.92 | 9.96 | 3.09 | 0.065 | 0.83 | 0.60377 |
| | FC-OE | 7.69 | 2.78 | 6.9 | 2.59 | 0.2544 | **7.5** | **2.66** | **15.38** | **3.63** | **0.0059** | 0.36 | 0.05575 |
| | FC-CE | **13.52** | **5.5** | **10.71** | **1.34** | **0.0762** | 15.24 | 4.11 | 27.1 | 7.21 | 0.0329 | **0.9** | **0.02532** |
| **RMS** | SP-OE | 77.42 | 12.43 | 55.46 | 10.81 | 0.1842 | 96.75 | 20.16 | 221.64 | 33.45 | 0.1756 | 0.66 | 0.2305 |
| | SP-CE | 104.21 | 14.91 | 77.7 | 13.43 | 0.1864 | 74.73 | 12.25 | 256.27 | 44.21 | 0.0874 | -0.38 | 0.40721 |
| | FC-OE | 103.34 | 16.76 | 79.75 | 11.67 | 0.2312 | 101.26 | 14.36 | 494.37 | 57.17 | 0.0275 | 0.59 | 0.10912 |
| | FC-CE | 348.7 | 73.13 | 183.41 | 34.43 | 0.0626 | **739.65** | **173.64** | **1561.53** | **105** | **0.0451** | -0.35 | **0.04117** |
| **Path** | SP-OE | **537.72** | **109.1** | **333.67** | **66.85** | **0.00136** | 2315.5 | 592.94 | 2283.21 | 693.8 | 0.46877 | 0.42 | **0.0000** |
| | SP-CE | **947.38** | **212.76** | **505.68** | **124.11** | **0.00286** | 2727.53 | 737.13 | 3264.01 | 672.93 | 0.24802 | 0.8 | **0.0000** |
| | FC-OE | **1012.91** | **316.34** | **515.48** | **107.25** | **0.0001** | 3084.45 | 570.2 | 3831.15 | 488.17 | 0.19274 | 0.05 | **0.0000** |
| | FC-CE | **2842.47** | **759.51** | **1409.81** | **583.18** | **0.00006** | 8252.69 | 1496.54 | 7148.73 | 1276.84 | 0.31056 | 0.36 | **0.0000** |
| **Range** | SP-OE | 42.91 | 9.64 | 34.24 | 8.73 | 0.1085 | 54.58 | 13.01 | 76.59 | 18.87 | 0.1425 | **0.76** | **0.02559** |
| | SP-CE | 47.96 | 11.19 | 41.78 | 7.9 | 0.1642 | 48.15 | 10.27 | 110.37 | 27.05 | 0.0475 | 0.78 | 0.13619 |
| | FC-OE | 53.58 | 12.12 | 42.22 | 10.64 | 0.0745 | **55.47** | **14.61** | **147.87** | **20.42** | **0.0066** | **0.79** | **0.02808** |
| | FC-CE | **101.91** | **22.45** | **68.52** | **7.88** | **0.0156** | 181.25 | 30.6 | 205.45 | 44.96 | 0.4128 | 0.97 | 0.05762 |
| **MF** | SP-OE | 0.11 | 0.04 | 0.09 | 0.04 | 0.1657 | 0.5 | 0.14 | 0.5 | 0.18 | 0.4976 | **-0.93** | **0.0000** |
| | SP-CE | **0.17** | **0.04** | **0.1** | **0.03** | **0.0007** | **0.55** | **0.1** | **0.44** | **0.11** | **0.0483** | 0.65 | **0.0000** |
| | FC-OE | **0.18** | **0.03** | **0.1** | **0.03** | **0.0001** | **0.61** | **0.19** | **0.34** | **0.15** | **0.0012** | -0.53 | **0.0000** |
| | FC-CE | **0.29** | **0.07** | **0.17** | **0.08** | **0.0012** | **0.71** | **0.13** | **0.45** | **0.12** | **0.0089** | -0.12 | **0.0000** |
| **Area** | SP-OE | 24236.25 | 5180.33 | 9909.11 | 2260.59 | 0.0045 | 56187.86 | 9376.84 | 154228.5 | 21038.88 | 0.0459 | 0.07 | 0.01874 |
| | SP-CE | 35693.62 | 7323.77 | 14694.15 | 2405.05 | 0.0103 | 49174.5 | 6898.21 | 455780.41 | 45459.84 | 0.0504 | 0.37 | 0.012886 |
| | FC-OE | 47539.11 | 6921.1 | 21722.47 | 4405.45 | 0.0025 | 70412.12 | 10142.29 | 776658.33 | 71616.59 | 0.0178 | 0.11 | 0.06251 |
| | FC-CE | 166576 | 23829.34 | 50165.57 | 10159.06 | 0.0006 | 433879.8 | 54227.5 | 1391118.98 | 98335.51 | 0.0411 | 0.36 | 0.0247 |
| **PWR** | SP-OE | **19.27** | **6.34** | **9.62** | **5.78** | **0.0149** | 21.24 | 9.48 | 125.51 | 21.94 | 0.0466 | 0.35 | 0.11985 |
| | SP-CE | **35.02** | **11.44** | **15.38** | **6.36** | **0.0211** | 28.57 | 10.89 | 703.77 | 135.01 | 0.0891 | 0.31 | 0.26012 |
| | FC-OE | **42.62** | **15.66** | **19.75** | **8.68** | **0.0249** | 31.92 | 12.75 | 900.58 | 191.91 | 0.0436 | 0.82 | 0.1641 |
| | FC-CE | **142.67** | **33.4** | **50.15** | **13.78** | **0.0126** | 355.32 | 73.9 | 1209.77 | 144.64 | 0.0981 | 0.94 | 0.06832 |
| **F50** | SP-OE | 0.19 | 0.01 | 0.18 | 0.00 | 0.0612 | 0.28 | 0.05 | 0.31 | 0.14 | 0.2707 | **0.27** | **0.00001** |
| | SP-CE | 0.21 | 0.02 | 0.2 | 0.01 | 0.4553 | **0.29** | **0.06** | **0.25** | **0.05** | **0.0465** | 0.93 | **0.00003** |
| | FC-OE | 0.21 | 0.02 | 0.2 | 0.04 | 0.3923 | **0.35** | **0.06** | **0.27** | **0.01** | **0.0161** | 0.25 | **0.0000** |
| | FC-CE | 0.27 | 0.04 | 0.27 | 0.06 | 0.4073 | **0.35** | **0.06** | **0.27** | **0.01** | **0.0161** | 0.62 | **0.04192** |
| **F95** | SP-OE | 0.99 | 0.2 | 0.88 | 0.21 | 0.1174 | 2.19 | 0.38 | 1.92 | 0.42 | 0.1402 | **0.91** | **0.0000** |
| | SP-CE | 0.78 | 0.22 | 0.89 | 0.15 | 0.1113 | 1.89 | 0.42 | 1.9 | 0.53 | 0.4858 | **0.93** | **0.0000** |
| | FC-OE | 0.85 | 0.23 | 0.93 | 0.08 | 0.1772 | **2.34** | **0.47** | **1.84** | **0.6** | **0.0391** | 0.94 | **0.0000** |
| | FC-CE | 0.84 | 0.19 | 0.76 | 0.08 | 0.1324 | **2.77** | **0.57** | **1.74** | **0.6** | **0.0421** | 0.43 | **0.0000** |
| **CF** | SP-OE | 0.32 | 0.03 | 0.3 | 0.03 | 0.1414 | 0.53 | 0.1 | 0.55 | 0.15 | 0.4018 | -0.37 | **0.0000** |
| | SP-CE | 0.29 | 0.04 | 0.31 | 0.02 | 0.1464 | 0.51 | 0.08 | 0.49 | 0.13 | 0.31 | -0.83 | **0.0000** |
| | FC-OE | 0.31 | 0.04 | 0.33 | 0.03 | 0.1205 | **0.59** | **0.13** | **0.48** | **0.16** | **0.0394** | 0.54 | **0.0000** |
| | FC-CE | 0.34 | 0.05 | 0.34 | 0.04 | 0.3525 | **0.64** | **0.1** | **0.51** | **0.16** | **0.0213** | 0.86 | **0.0000** |
| **FD** | SP-OE | **0.5** | **0.12** | **0.25** | **0.11** | **0.0207** | 0.34 | 0.19 | 2.2 | 0.66 | 0.0411 | -0.01 | 0.19063 |
| | SP-CE | **0.99** | **0.21** | **0.38** | **0.15** | **0.0187** | 0.36 | 0.17 | 12.93 | 3.3 | 0.09 | -0.9 | 0.30152 |
| | FC-OE | 1.28 | 0.6 | 0.48 | 0.1 | 0.08 | **0.36** | **0.19** | **17.25** | **3.29** | **0.0395** | -0.13 | 0.19055 |
| | FC-CE | 3.67 | 1.11 | 0.97 | 0.39 | 0.0661 | 2.99 | 1.04 | 19.96 | 6.65 | 0.058 | -0.73 | 0.17618 |

*In boldface are reported significant different measures between unilateral vestibular failure (UVF) and healthy subjects (HS) and significant correlations between accelerometer (ACC) and centre of pressure (COP) parameters. standard deviation, SD; mean distance from centre of COP (ACC) trajectory, DIST; root mean square of COP (ACC) time series, RMS; total length of COP (ACC) trajectory, PATH; range of COP displacement (acceleration), RANGE; mean frequency, the number, per second, of loops that have to be run by the COP (ACC), to cover a total trajectory equal to PATH, MF; sway area, AREA; total power, PWR; median frequency, F50; 95% power frequency, F95; centroidal frequency, CF; frequency dispersion ,FD; eyes closed, CE; eyes opened, OE; stiff platform , SP; foam carpet, FC.*

in discriminating UVF from healthy subjects. Interestingly, on one hand they are in line with the natural history of these kind of vestibular disorders in which length and surface values were addressed to be higher than healthy subjects [3]. On the other, a new interesting ACC-based and length-derived parameter (see Table I), such as MF, could be proposed as a further measure to follow UVF patients during the course of the disease. Finally, two frequency-domain measures (F50 and F95) were also demonstrated to be reliable parameters in discriminating UVF subjects when assessed by ACC recordings, especially in situations in which the balance integration system is involved in more difficult sensory input conditions (i.e. FC and CE; see Table II). This could be of interest for clinicians, and in particular when the vestibular system will be studied – under frequency-domain parameters – to assess its possible relationships with risk of fall.

These overall aspects should further highlight the attention of clinicians and researchers on this type of sway recording technique in the field of otoneurological disorders, considering the possibility to enrich the amount of quantitative and qualitative information useful in discrimination, diagnosis and treatment of UVF. In addition, the ACC body sway recording provides a large number of measures that automatically, fully characterise body sway in amplitude, smoothness and frequency; these measures are relevant for testing any individual with balance deficits. Thus, for future perspectives its potential application should not be limited to testing subjects with UVF.

In fact, it is likely that a different subset of ACC-based measures might be sensitive to different constraints. Further studies are needed to determine the best subset of postural sway parameters that can predict body sway or stance disability during daily activities in many neuro-otological, neurological, systemic, or musculoskeletal disorders.

Finally, the ease distinguishing this kind of technique can supply further information in case of multiple body-worn sensors worn on patients and in those conditions in which the recording trace is investigated during patient-tailored tasks or sensory feedback.

## Conclusions

The present accelerometric-based measurement of sway offers a patient-friendly, reliable, inexpensive and efficient alternative recording technique that is useful to quantify and qualify posture control. This method demonstrated – under controlled conditions – a high level of reliability compared to gold-standard FBP measures. Thus, it could be used together with clinical balance and mobility tests in various circumstances, particularly in outcome studies, involving diagnosis, follow-up and rehabilitation of UVF patients.

## Funding

The research was partially funded by the European Union's Horizon 2020 research and innovation programme under the Marie Sklodowska-Curie grant agreement No 659434.

Address for correspondence: Alessandro Micarelli, Department of Clinical Sciences and Translational Medicine, Tor Vergata University, via Montpellier, 1, E sud Tower, 00133 Rome, Italy. Tel. +39 06 20902925. Fax +39 06 20902930. E-mail: alessandromicarelli@yahoo.it